\documentclass[useAMS,usenatbib]{mn2e}
\usepackage{epsfig}                     
\usepackage{amssymb}
\usepackage{color}
\usepackage{multicol}
\title[Thermonuclear double-detonations as possible SNe Ia]{Delay Times and Rates for
Type Ia Supernovae and Thermonuclear Explosions from Double-detonation
Sub-Chandrasekhar Mass Models}
\author[A.J.Ruiter et al.]
{\parbox{\textwidth}{
A.J.~Ruiter,$^{1}$\thanks{E-mail: \texttt{ajr@mpa-garching.mpg.de; kbelczyn@astrouw.edu.pl; ssim@mso.anu.edu.au;
wfh@mpa-garching.mpg.de; fryer@lanl.gov; mfink@mpa-garching.mpg.de;
mkromer@mpa-garching.mpg.de}}
K.~Belczynski,$^{2,3}$
S.A.~Sim,$^{1,4}$
W.~Hillebrandt,$^{1}$
C.L.~Fryer,$^{5}$
M.~Fink,$^{1}$ and 
M.~Kromer$^{1}$}\vspace{0.4cm}\\
\parbox{\textwidth}{$^{1}$ Max Planck Institute for Astrophysics,
Karl-Schwarzschild-Str. 1, 85741 Garching, Germany\\
$^{2}$ Astronomical Observatory, University of Warsaw, Al.
            Ujazdowskie 4, 00-478 Warsaw, Poland\\
$^{3}$ Center for Gravitational Wave Astronomy, University of Texas at
            Brownsville, Brownsville, TX 78520, USA\\
$^{4}$ Research School of Astronomy and Astrophysics, Mount Stromlo
Observatory, Cotter Road, Weston Creek, ACT 2611, Australia \\
$^{5}$ Los Alamos National Laboratory, CCS-2, MS D409, Los Alamos, NM 87545 USA}}

\begin{document}

\date{Accepted 2011 June 16. Received 2011 June 16; in original form
  2010 November 03}

\pagerange{\pageref{firstpage}--\pageref{lastpage}} \pubyear{2010}

\maketitle

\label{firstpage}

\begin{abstract}

  We present theoretical delay times and rates of thermonuclear explosions that 
  are thought to produce Type Ia supernovae, 
  including the double-detonation sub-Chandrasekhar mass model, using the
  population synthesis binary evolution code {\tt StarTrack}. 
  If detonations of sub-Chandrasekhar mass carbon-oxygen
    white dwarfs following a detonation in an accumulated layer of helium
 on the white dwarf's surface (``double-detonation'' models) are able to
 produce thermonuclear explosions which are characteristically similar 
 to those of SNe Ia, then these sub-Chandrasekhar mass explosions may 
 account for at least some substantial fraction of the observed SN Ia
 rate. Regardless of whether all double-detonations look like
 `normal' SNe Ia, in any case the explosions are expected to be 
 bright and thus potentially detectable.  
 Additionally, we
  find that the delay time distribution of double-detonation 
  sub-Chandrasekhar mass SNe Ia can be divided into two distinct
  formation channels: the `prompt' helium-star channel with delay
  times ${<}500$~Myr ($13\%$ of all sub-Chandras), and the `delayed' 
  double white dwarf channel, with delay times ${\gtrsim}800$~Myr
  spanning up to a Hubble time ($87\%$).  These findings coincide with recent
  observationally-derived delay time distributions which have revealed
  that a large number of SNe Ia are prompt with delay times
    ${<}500$~Myr, while a significant fraction also have delay
    times spanning $\sim 1$~Gyr to a Hubble time.

\end{abstract}


\section{Introduction}

The exact nature of the stars that produce Type Ia supernovae (SNe Ia)
-- which are believed to be thermonuclear explosions of carbon--oxygen
(CO) white dwarfs (WDs) close to the Chandrasekhar mass limit --
remains unknown \citep[e.g.,][]{Bra95}.  The most widely favoured SN Ia progenitor scenarios
involve the double degenerate scenario \citep[DDS;][]{Web84,IT84}, and the
single degenerate scenario \citep[SDS;][]{WI73}.  In the DDS, the
merger of two CO WDs with a total mass exceeding the Chandrasekhar
mass limit, $M_{\rm Ch} \sim 1.4~M_{\odot}$, can lead to explosive
carbon burning which causes a SN Ia explosion.  In the SDS, a CO white
dwarf accretes from a hydrogen-rich stellar companion via stable
Roche-Lobe overflow (RLOF) and undergoes hydrogen burning on the
surface, enabling the WD to accumulate mass toward $M_{\rm Ch}$ until
carbon is ignited explosively in the centre of the WD leading to a SN
Ia.  However, in the stable RLOF configuration the companion does not
have to be a hydrogen-rich main sequence (MS) or giant-like star, but 
can be a non- or semi-degenerate helium-burning star, or a (degenerate) 
helium white dwarf \citep[e.g.,][]{Ibe87,YL03,SY05}.  
Like the SDS, the WD explodes once it approaches $M_{\rm Ch}$ (this helium-rich donor 
scenario will hereafter be referred to as HeRS).

Recently, Ruiter et al. 2009 (Paper~I) carried out a population
synthesis study showing rates and delay times -- time from birth of a
progenitor system in a short burst of star formation to SN -- for
three formation channels of SNe Ia: DDS, SDS and HeRS. 
Additionally it is worth asking the following question: to what degree do 
{\em sub-Chandrasekhar mass} WDs contribute to the population of explosive 
(SN Ia-like) events?  We point out that in some scenarios, a layer of accumulated 
helium on an accreting, sub-Chandrasekhar mass WD's surface 
may undergo several shell flashes \citep[e.g., leading up to a ``.Ia'' as discussed in][]{Bil07}, 
and while these explosions can be bright, none of them result in a SN Ia.  However, it is possible 
that the detonation in the helium layer causes the underlying CO WD to detonate, resulting 
in a (final) explosion that does look like a SN Ia.
Here, we extend our study of progenitors and focus our
analysis and discussion on these thermonuclear ``double-detonation''
events involving helium-rich donors, hereafter referred to as the sub-$M_{\rm Ch}$
model, in which a CO WD accretes from a helium-rich companion filling its
Roche-Lobe and explodes as a SN Ia before reaching the $M_{\rm Ch}$
limit.  

The sub-$M_{\rm Ch}$ model 
has thus far been regarded as an unlikely model for SNe Ia
owing to the fact that most synthetic light curves and spectra of
these objects from previous studies did not match those observed for
SNe~Ia.  However, it has recently been argued that the
model might be capable of producing a better match to observation,
depending on details regarding the manner in which the accreted helium
burns \citep[e.g.][]{Fin10,Kro10}. In either case, the explosion
mechanism is expected to produce events that are bright and should be
detectable.  Thus, quantification of their predicted rates and delay
times is an important step for testing our population synthesis
models, and for determining what fraction of SNe~Ia could conceivably
be associated with this channel.

Since these calculations are based on the work that was performed for
Paper~I, the reader is referred to that paper for a more detailed
description of the DDS, SDS and HeRS 
scenario.  The layout of this paper is as follows: In section 2 we
summarise some background information on SN Ia progenitors from the
literature.  In section 3 we discuss the population synthesis
modelling.  In section 4 we present delay time distributions and rates
as a function of stellar mass as well as distributions showing the
exploding CO WD core mass.  In section~5 we close with a
discussion of these findings and possible implications/predictions for
SN progenitors and their host stellar populations.

\section{Background}
\subsection{Recent observations of SNe Ia delay times}

The idea that SN Ia progenitors belong to at least two distinct
populations \citep[][]{SB05,MDP06,PHS08} has been
gaining ground.  A picture is emerging
which supports populations of both quickly-evolving (prompt)
progenitors with short delay times less than ${\sim}500$~Myr, as well
as more slowly-evolving progenitors with (sometimes rather) long delay
times spanning up to a Hubble time \citep[but see also][]{Gre10}.

The delay time
distribution (DTD) is a useful tool in determining the age of the
progenitor stellar population, which places strong constraints on the
different proposed progenitor scenarios.  There are a growing number
of observationally-derived DTDs presented in the literature from
various groups \citep[see section 1 of][for an overview of these
previous studies]{MB10}.  
\citet{Tot08} derived
the DTD from a large population of old galaxies which only probed
delay times ${\gtrsim}100$~Myr, and they found that the DTD follows a
relatively smooth power-law distribution ($t^{-1}$) from ${\sim}0.1$
to $8$~Gyr \citep[see also][]{HB10}. 
Probing younger stellar populations,
\citet{MB10} were able to determine that a substantial fraction of SNe Ia 
are prompt.\footnote{A
  complimentary result was also determined by \citet{MSG10}, who found
  that in galaxy clusters the DTD is
  well-fit by a power-law of $t^{-1.2}$ for delay times $>400$~Myr.}  
Among these prompt SNe Ia ($35$--$330$~Myr delay times in that study) the
SN Ia rate\footnote{The supernova rate in the local Universe as a
  function of Hubble type was recently presented in
  \citet{Li10a} who found the SN Ia rate to be constant across
  galaxy Hubble type, with a value of $0.136 \pm 0.018$~SNuM (1 SNuM = SNe Ia
  per century per $10^{10}~M_{\odot}$).} is
${\sim}0.09$--$0.40$~SNuM, compatible with the results of
\citet{Li10a}, whereas delayed SNe Ia in that study ($330$~Myr--$14$~Gyr delay
times) had an overall smaller rate: ${<}0.0024$~SNuM\@.  This study
confirmed that roughly half of SNe Ia
occur with delay times ${\lesssim}330$~Myr, thus giving strong
support for a prompt component of the DTD. 
\citet{Mao10}
  reconstructed the star formation histories for a sample of LOSS SN
  host galaxies and found strong evidence for both a prompt Ia component with 
  delay times $< 420$ Myr and a delayed component with long delay times ($>2.4 $
Gyr).  \citet{Bra10} used SN
light-curves and spectra from host galaxies of 101 SNe Ia with $z<0.3$
to construct the DTD, and arrived at a similar conclusion: that
roughly half of SNe Ia occur with delay times ${<}400$~Myr, while
the other half have long ($>2.4$~Gyr) delay
times.  Further, they find that the short delay time events are more
luminous with slowly-declining light-curves, and are associated with
young stellar populations, whereas the SNe with long delay times are
typically fast-declining, sub-luminous events.  

As one can see, the  aforementioned recent studies indicate that SNe Ia are observed to occur 
over a range of long delay times (with less events as time goes on), with a 
substantial fraction also occurring at very early times.  

\subsection{Two progenitor scenarios: DDS and SDS}
 
For some time, population synthesis calculations
\citep{IT84,Yun94,YT97,YL98,Nel01b,RBF09}
have predicted that the number of merging CO WDs with a
total mass exceeding $M_{\rm Ch}$ (DDS) is sufficient to match, and thus
possibly account for, the rate of SNe Ia \citep[$0.4 \pm 0.2$ per
century for the Galaxy,][]{CET99}.  At the same time, 
the theoretically-predicted SN Ia rate from the SDS channel is usually
unable to explain the observed rates of SNe Ia \citep[see
also][]{GB10}.  
There are very few SNe Ia that show any hint of hydrogen lines in their
spectra; if the progenitor involved a hydrogen-rich companion, in
  particular a giant donor, one may
expect H$\alpha$ to be detectable in the nebular spectra more
frequently \citep{Leo07,Hay10}. 

The DDS is an attractive model for SNe
Ia, given the theoretically predicted occurrence rate as well as the
fact that CO WD mergers are systems that are essentially devoid of
hydrogen.  
The main argument against the DDS is that detailed WD merger
calculations between two CO WDs with a total mass ${>} M_{\rm Ch}$
indicate that the merging process, while it can 
potentially lead to a thermonuclear explosion if the correct 
conditions are satisfied \citep[][]{YPR07}, 
is more likely to result in collapse and form a neutron
star; an accretion induced collapse \citep[AIC,][and references
therein]{Miy80,SN85,NK91}. 
Based on modern collapse calculations~\citep{Fry99,Des07,Abd10},
\citet{Fry09} found that these AICs produced outbursts that were
${\sim}1$--3 magnitudes dimmer than typical SNe Ia, arguing
that AICs could only explain a few abnormal Ia explosions.

In a merger of two CO WDs, once the larger (less massive) WD fills its
Roche-lobe, it is likely to be disrupted and rapidly accreted by the
companion.  This process can be quite violent, and might under the
right conditions lead to a SN Ia explosion \citep{Pie03}.  For
example, in the DDS case involving the merger of two WDs with
a mass ratio close to unity and WD masses 
${\sim} 0.9~M_{\odot}$ \citep{Pak10}, critical conditions for the
successful initiation of a detonation \citep{Sei09} can be obtained.
The \citet{Pak10} study found that these DDS systems can both in
number and in observational characteristics account for the population
of sub-luminous 1991bg-like SNe Ia.  However, the 1991bg-like systems
only account for a small fraction of SNe Ia \citep{Li10b}.

For lower mass ratios \citep[see][Pakmor et al. 2011]{Pak10} it
is unlikely that the merger would lead to a SN Ia as the achieved
densities are not high enough to enable a detonation to occur.  In
such WD mergers, high accretion rates onto the relatively `cold'
primary WD can lead to carbon burning off-centre, where the densities
are too low, and carbon does not burn explosively.  The primary CO WD
will in turn burn carbon and evolve into an oxygen--neon--magnesium
(ONeMg) WD \citep{NK91}.  As the ONeMg WD increases in mass, density
and temperature conditions become more favourable for electron
captures, which in turn remove electron degeneracy pressure from the
undisrupted WD\@.  As the WD approaches $M_{\rm Ch}$, the central
densities continue to increase and the WD collapses to become a
neutron star before a thermonuclear explosion can take place.

Despite this, in population synthesis calculations it is typically assumed that all
mergers of CO--CO WDs produce a SN Ia provided that the total mass
exceeds $M_{\rm Ch}$.  If some (or many) of these CO--CO mergers
result in AIC then the observed SN Ia rates cannot be fully explained
by the DDS model. Thus, if it is true that the majority of the DDS
systems cannot produce events that look like SNe Ia, and there are
not enough SDS or HeRS events, then a
significant fraction of SNe Ia remain to be accounted for.

\subsection{The helium donor formation channels}

\subsubsection{Chandrasekhar mass explosions}

We delineate between the SDS and HeRS 
since the latter can involve either one degenerate star 
where the donor is helium-burning, or two, where the donor is a
helium-rich WD.  Double WDs will 
have very close orbits (orbital periods ${<} 70$~minutes) 
as is the case for typical AM CVn
binaries \citep[see][for a discussion on AM CVn stars]{Nel01a,Nel10}.
In Paper~I the rates and delay times from three formation channels
involving exploding WDs with masses $\ge M_{\rm Ch}$ (DDS, SDS and
HeRS) were investigated.  Both helium-donor
channels were referred to as the `AM CVn channel' in
Paper~I.  
In this paper we adopt the acronym HeRS 
for all SN Ia progenitors in which the
{\em Chandrasekhar mass WD} explodes once it has accreted sufficient mass in
stable RLOF from a helium-rich companion, whether the donor is degenerate 
or non-degenerate.  This scenario includes AM CVn binaries as well as WDs accreting from
all helium-burning stars. 

SN Ia rates of the HeRS scenario leading to
SNe Ia have been previously investigated by different groups:
\citet[][]{SY05,RBF09,Wan09a,Wan09b,MY10}, some of whom considered
only the helium-burning star channel.  
In the majority of studies it
was found that the HeRS is unable to account for the rates of SNe
Ia.\footnote{\citet{Wan09a}
found Galactic SN Ia rates which were higher than the other studies:
${\sim} 10^{-3} \, \mathrm{yr}^{-1}$.  However this rate is
likely somewhat optimistic because they consider a rather
large range of orbital periods at the moment of RLOF onset between the
WD and the helium star (up to ${>} 100$~days for the most massive WD
accretors).}  In many cases, theoretically-motivated studies of the
HeRS channel produce SNe Ia with short delay
times (${\lesssim}$ a few hundred Myr), and are not able to account for a
large number of systems at long delay times. 

\subsubsection{sub-Chandrasekhar mass explosions}

Thermonuclear explosions may occur in systems with a sub-$M_{\rm Ch}$
(probably CO) WD accreting via stable RLOF from a helium-rich
companion \citep{IT91,TY96,YL00}. 
It has been shown that at certain (low) accretion
rates on to the WD helium flashes on the WD surface are inhibited  
\citep{KSN87,IT04}, and the WD can
steadily and efficiently build up a massive layer of helium on the
surface.  In such a massive degenerate helium shell \citep[e.g.,
${\sim} 0.1~M_{\odot}$ of helium,][]{Taa80}, a flash may likely evolve
as a violent detonation, which may also trigger a detonation of the CO
core, and thus a thermonuclear explosion of the complete star
\citep[e.g.,][]{Liv90,WW94,LA95}.\footnote{It is possible that the companion can
  be hydrogen-rich 
  where hydrogen burns steadily on the surface
  of the WD, building up a helium-rich layer on top of the CO WD which
  can detonate \citep{Ken93,Pie99,YL00}. Such a double-detonation
    scenario for hydrogen-rich donors was investigated by \citet{Yun95}, who have shown
  that sub-Chandrasekhar mass WDs accreting hydrogen
  in symbiotic binaries may be capable of producing 
  up to $\sim \frac{1}{3}$ of all SNe Ia, provided that 
  accreting WDs with masses as low as $0.6~M_{\odot}$ are able to 
  successfully undergo double-detonations.  However in this work, we
  only consider helium-rich donors as possible companions for
  double-detonation sub-$M_{\rm Ch}$ SN Ia
  progenitors.}
\citet{TY96} found that the rate of sub-$M_{\rm Ch}$ SNe Ia 
from the non-degenerate helium star channel might be high enough to 
account for the Galactic rate if these explosions 
are comparable in luminositiy to normal SNe Ia, though these 
types of events were found to have somewhat short delay times 
and cannot account for the number of SNe Ia in old stellar 
populations.

Sub-$M_{\rm Ch}$ models of SNe~Ia are appealing for a number of
reasons. 
Population synthesis calculations have already shown that
the DTD for the double-detonation sub-$M_{\rm Ch}$ explosion 
model \citep[sometimes referred to as edge-lit detonation, ELD; see][and references therein]{YL00} spans a wide range at early times.
\citet{YL00} investigated the double-detonation scenario for 
helium star and hydrogen-rich donors, and found a corresponding 
delay time of $\sim 30$ Myr$- 1.3$ Gyr for the helium star 
donor systems.  Also, population synthesis modelling has indicated 
that the number of potential progenitors for the double-detonation
sub-$M_{\rm Ch}$ SNe Ia involving helium-rich donors alone may be large 
enough to account for the observed rates
\citep[e.g.,][]{Tou01,Reg03,RBF09}.  Additionally, 
simplistic studies of pure detonations of sub-$M_{\rm Ch}$ WDs (in the
absence of any overlying helium shell) indicate that the synthetic
light curves and spectra from sub-$M_{\rm Ch}$ explosions may be able
to reproduce a surprising number of the observed properties of SNe~Ia
\citep{Sim10}, and even variations within the class that could be
associated with differences in the mass of the exploding WD\@.

The \citet{Sim10} work, however, neglects the issue of how the outer
helium layer will affect the observables.  Several previous studies
have calculated detailed synthetic light-curves and spectra of
sub-$M_{\rm Ch}$ double-detonation models
\citep[e.g.][]{HK96,Nug97}. These concluded that such explosions would
likely not lead to events with observational properties characteristic
of normal SNe Ia. In general, the light curves were found to rise and
fall too rapidly compared to `normal' SNe~Ia while their spectra were
too blue to match sub-luminous SNe~Ia and lacked sufficiently strong
features of intermediate mass elements, such as Si and S.
Importantly, most of these discrepancies with observation can be
traced to the presence of the products of helium burning ($^{56}$Ni
and other iron-group elements) in the outer regions of the ejecta, and
those studies mainly considered systems in which a relatively massive
(${\sim} 0.2~M_{\odot}$) helium layer had accumulated on the WD
(${\sim} 0.6~M_{\odot}$) surface.
More recently, \citet{Bil07} and \citet{SB09} have shown that
conditions suitable for detonation in the WD might be reached for
somewhat lower helium shell masses than considered in most previous
studies: perhaps as low as $0.05~M_{\odot}$ for a CO WD (core) mass of
$1.0~M_{\odot}$ (in general the more massive the CO WD, the less
massive the accumulated helium layer needs to be for a detonation).
\citet{Fin10} have shown that, even for such low helium shell masses,
detonation of the helium will robustly lead to an explosion of the
underlying WD\@.  With a significantly lower He shell mass (and thus
fewer iron-group elements in the outer ejecta), this may open the door
for double-detonation sub-$M_{\rm Ch}$ models whose spectra and light
curves are in better agreement with observed SNe~Ia.
This has been investigated by \citet{Kro10} who computed synthetic
observables for the \citet{Fin10} simulations. They showed that even
very low mass ($0.05~M_{\odot}$) helium shells affect the observable
display and can lead to spectroscopic signatures that are not
characteristic of observed SNe Ia. However, \citet{Kro10} also
highlighted that the results are highly sensitive to the details of
the nucleosynthesis that occur during burning of the helium shell.
Modifications to the burning -- as might be achieved by considering a
composition other than pure helium -- could allow the model
predictions to achieve much better agreement with
observation.

Taken together, the body of theoretical work strongly suggests that
the sub-$M_{\rm Ch}$ double-detonation scenario is physically
realistic.  Depending on the details of the accumulated helium layer
and its burning products, the explosion may closely resemble observed
``normal'' SNe~Ia or it might be highly spectroscopically peculiar --
but regardless of this, it certainly can be bright enough to be
readily observable -- for a CO WD of around 1.0~$M_{\odot}$ the
luminosity produced following detonation is expected to be close to
that of a normal SN~Ia \citep{Shi92,Sim10}.  Given that potential
progenitors are also expected to be common, we are therefore compelled
to further investigate this progenitor scenario. If these explosions
can produce events that resemble ``normal'' SNe~Ia then it is of
interest to quantify the fraction of observed SNe~Ia that might be
accounted for via this channel. Alternatively, if these explosions are
realised in nature but are spectroscopically peculiar, it is important
to estimate their predicted rate and consider whether the apparent
lack of observational detections is a major concern for the
established theory; the lack of such events may challenge our
understanding of either (or both) the explosion physics or the
progenitor binary evolution. For this scenario it is of particular
interest to consider the DTD predicted for this class of explosion and
to investigate any correlations between the properties of the
exploding system (particularly the mass of the primary WD, which
likely determines the brightness of the explosion) and the age of the
stellar population in which it resides.

\section[]{Model}

It has been shown that a WD accumulating
helium-rich material may be capable of exploding as a SN Ia if the
correct conditions are satisfied, even if the WD is below
$M_{\rm Ch}$ \citep{Taa80,Ibe87,IT91,WW94,LA95,IT04}.  In \citet{BBR05},
sub-$M_{\rm Ch}$ models were calculated, though DTDs
for different formation channels were not discussed
separately, and rates were not presented in that work.  In Paper~I,
rates and delay times were presented only for ${\gtrsim} M_{\rm Ch}$
WD mass models where as here, we additionally include
sub-$M_{\rm Ch}$ SNe Ia progenitors in our study.

All sub-$M_{\rm Ch}$ SN progenitors in our
calculations involve a CO WD accreting via RLOF from a helium-rich companion.
As in Paper~I, if the donor is a WD then it can be either a 
helium WD or a hybrid WD; a WD with a CO core and a helium-rich
mantle \citep[e.g.,][and references therein]{TY96}.  Hybrid WDs are
formed through binary evolution when a red giant is stripped 
of its envelope through binary interactions.  
In cases where the stripped helium core does not reach
the helium-burning phase, a helium WD is formed.

In the following sections we compute and discuss rates and delay 
times for the aforementioned SN Ia evolutionary models that have been 
proposed as the most promising formation channels for SNe Ia:
\begin{itemize}
\item{the DDS}
\item{the SDS}
\item{the HeRS}
\item{the double-detonation sub-$M_{\rm Ch}$ scenario involving
    helium-rich donors}
\end{itemize}

It is important to keep in mind that, while evolution of close
binaries remains an active field of research and discovery, no
concrete constraints currently exist for the evolution of
mass-transferring binaries, nor for the common envelope (CE) phase.
The CE phase is certainly one of the most
poorly understood phenomena in close binary evolution, and a 
theoretical picture of CE evolution is not yet available.  Since there
are a limited (though growing) number of observations available to
guide our choice of parameters, we present results for three different
common envelope realizations which most effectively bracket the
uncertainties.  For the growth of CO WDs during stable RLOF, as was
done in \citet{RBF09} we present the results from our population
synthesis model using a detailed WD accretion scheme, which was
constructed by adopting various input physics from the literature.

We use the {\tt StarTrack} population synthesis binary evolution code
\citep{Bel08} to evolve our stellar populations employing Monte Carlo
methods.  The code has undergone many revisions since the first code
description publication \citep{BKB02}.  Many of the updates concerning
accretion on to WDs can be found in \citet{BBR05,Bel08}, though
since then we have incorporated an updated prescription for accretion
of hydrogen on WDs by including calculations from \citet{Nom07} 
in addition to \citet{PK95}.  The initial distributions for
binary orbital parameters (orbital periods, mass ratios, etc.)  are
the same as described in Paper~I, section 2.

In Paper~I, it was assumed that the ejection of the envelope of the
mass-losing star during a CE phase came at the expense of removing the
orbital energy of the binary, as dictated by the well-known
`energy-balance' (or `$\alpha$-formalism') equation \citep{Web84}, with 
$\alpha_{\rm CE}$ representing the efficiency with which the binary orbital energy can
unbind the CE, and $\lambda$ is a parametrization of the structure of
the donor star \citep{deK90}; both $\alpha_{\rm CE}$ and $\lambda$ are
fairly uncertain.

For Models 1 and 2 from Paper~I, $\alpha_{\rm CE} \times \lambda$
values of $1$ and $0.5$ were adopted, respectively.  The major
difference was that Model 1 (more efficient removal of the CE)
resulted in an overall higher number of SNe. 
In the current paper, we keep all model parameters the same
as in Model 1 of Paper~I for one model; we refer to this model as
Model A1 (standard model).  However, in order to explore the
sensitivity of the physical mechanism of CE ejection, which is still
not understood, we have run two additional sets of models.  There has
been some recent observational \citep{Zor10} as well as theoretical
\citep[][in prep.]{Pas10} evidence that the value for $\alpha_{\rm CE}$ lies
between $0.2$ and $0.3$.  Additionally, for low-mass stars a value of
$0.5$ is often adopted for $\lambda$ \citep{vdS06}.  Thus to best
bracket our uncertainties for the energy balance prescription of CE
evolution, in a second model we employ a very low CE ejection
efficiency: $\alpha_{\rm CE} = 0.25$ and $\lambda = 0.5$ yielding
$\alpha_{\rm CE} \times \lambda = 0.125$.  We refer to this model as
Model A.125.  In a third model, we assume a different parametrization
for the treatment of the CE phase.  We employ the `$\gamma$'
prescription for CE evolution \citep{Nel00} every time a CE event is
encountered in the code. Although the combination `$\alpha\,\gamma$' 
prescription is considered to be the preferred prescription by some groups, 
the two formalisms stress different physics (one focusing on energy conservation, 
one on angular momentum balance), and we 
do not mix the formalisms in the current paper.
Henceforth we refer to the $\gamma$ model as Model
G1.5.
For Model G1.5, all physical parameters are identical to
Models A1 and A.125, except that when unstable mass transfer is
encountered and a CE ensues, the orbital separation of the binary
changes not as a consequence of removing gravitational binding
energy from the orbit, but linearly as a function of mass loss (and
hence angular momentum loss), parametrized by the factor $\gamma$:
\begin{equation}
  \frac {a_{\rm f}}{a_{\rm i}} =  \left( 1 - \gamma \, \frac{M_{\rm ej}}{M_{\rm tot,i}} \right)^{2} \, \frac {M_{\rm tot,f}}{M_{\rm tot,i}} \left( \frac {M_{\rm don,i} \, M_{\rm com}} {M_{\rm don,f} \, M_{\rm com}} \right)^{2}
\end{equation}
where $M_{\rm don,i}$ is the
initial mass of the (giant) donor star just prior to the CE, $M_{\rm ej}$ 
is the ejected mass (assumed to be the mass of the giant's
envelope), $M_{\rm com}$ is the mass of the companion (assumed to be
unchanged during the CE), $M_{\rm don,f}$ is the final mass of the
donor once the envelope has been ejected, $a_{\rm i}$ is the initial orbital separation, $a_{\rm f}$ is the final
orbital separation, and $M_{\rm tot,i}$ and $M_{\rm tot,f}$ represent the total mass of
the binary before and after CE, respectively.  Following \citet{Nel00},
we have chosen $\gamma = 1.5$.\footnote{We note here that the $\gamma$
  CE equation in \citet[][equation 55]{Bel08} is missing an exponent,
  though the CE evolution is properly carried out in the {\tt
    StarTrack} code with the correct equation.}

\subsection{Sub-Chandrasekhar mass model: Assumptions}

We adopt the prescription of
\citet{IT04}, applied to accretion from helium-rich companions only,
to determine when a particular binary undergoes a sub-$M_{\rm Ch}$ SN
Ia \citep[][see section 5.7.2 for equations]{Bel08}.  In short, we
consider three different accretion rate regimes for accumulation of
helium-rich material on \emph{all} CO WDs, adopting the input physics
for helium accretion on to WDs of \citet{KH99,KH04}.  
Clearly, our results are sensitive to the adopted helium accumulation efficiency.  
It may also be possible that for the less-common hybrid donors, the star's helium-rich 
envelope becomes stripped in RLOF before a helium 
shell detonation on the CO WD occurs.  This would have an effect on the purity of the 
helium-rich shell, though we do not investigate the consequences here.  We assume that mass 
is exchanged conservatively in all cases where the transferred material would be CO-rich.

At high
accretion rates ($\sim~10^{-6}$~$M_{\odot} \, \mathrm{yr}^{-1}$, 
and for all cases where the WD accretor mass is $<0.7~$M$_{\odot}$),
helium burning is stable and thus mass accumulation on the WD is fully
efficient ($\eta_{\rm acu} = 1$).  At somewhat lower accretion rates 
helium burning is unstable and the binary enters a helium-flash cycle,
thus accumulation is possible but is not fully efficient ($0<
\eta_{\rm acu} < 1$).  In both of these aforementioned accretion
regimes, the CO WD is allowed to accrete (and burn) helium, and its
total mass may reach $M_{\rm Ch}$ and explode as a SN Ia through the
HeRS channel.  However, for low accretion
rates ($\sim~10^{-8}$~$M_{\odot} \, \mathrm{yr}^{-1}$),
compressional heating at the base of the accreted helium layer plays
no significant role, and a layer of unburned helium can be accumulated
on the WD surface.  Following \citet{IT04}, we assume that if such a
CO WD accumulating helium enters this `low' accretion rate regime and
accumulates $0.1~M_{\odot}$ of helium on its surface, a detonation is
initiated at the base of the helium shell layer.  Consequently, a
detonation in the core of the CO WD is presumed to follow, and we
assume that a sub-$M_{\rm Ch}$ SN Ia takes place.
Only accreting WDs with a \emph{total mass} ${\geq} 0.9~M_{\odot}$ are
considered to lead to potential sub-$M_{\rm Ch}$ SNe Ia in this work,
since lower mass cores may not detonate, and if they do they are 
unlikely to produce enough radioactive nickel and hence will not be
visible as SNe Ia \citep[e.g.,][table 1]{Sim10}. Thus in all future
discussions we refer to sub-$M_{\rm Ch}$ systems whose total WD mass
(CO core + helium shell) is at least $0.9~M_{\odot} $ at the time of
SN Ia unless otherwise noted; for our population synthesis model, this
intrinsically implies that all exploding sub-$M_{\rm Ch}$ SNe Ia have
CO WD `core' masses ${>} 0.8~M_{\odot}$.  Helium-rich WDs are simply
not massive enough, and we assume that ONeMg WDs do not make SNe Ia.

\section{Results}

Here we present the DTD and rates for all of our SN Ia
models, as well as CO WD core masses for our sub-$M_{\rm Ch}$ models.
Our results are discussed in the following subsections, but here we
give a brief outline of our findings:
 
We have investigated the DDS, SDS, HeRS, and
the sub-$M_{\rm Ch}$ scenario for three different CE realizations.
Within the framework of our adopted models, we find that only two SN
Ia formation scenarios are capable of matching the observed SNe Ia
rates: the DDS and the sub-$M_{\rm Ch}$ channels.  The most favourable
model in terms of matching observational rates is model A1 ($\alpha
\times \lambda = 1$).  For models A1 and G1.5 ($\gamma = 1.5$), the
adopted sub-$M_{\rm Ch}$ scenario %($M_{\rm WD} {\geq} 0.9~M_{\odot}$)
is dominant at nearly all epochs ${\lesssim} 5$~Gyr, however the
sub-$M_{\rm Ch}$ channel rate is too low for our low-efficiency CE
model (A.125).  For Model A.125, no single progenitor, nor an
admixture of all of the progenitors combined are able to account for
the observed rates of SNe Ia. 

\subsection{Delay times}

\begin{figure}
\includegraphics[width=\columnwidth,clip]{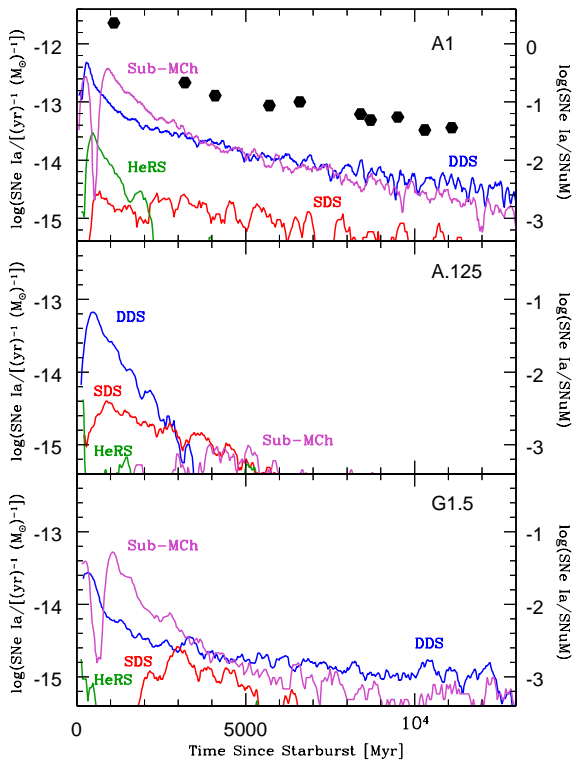}
\caption{Lines represent the DTD for SNe Ia. 
  Top panel: Model A1.  Middle
  panel: Model A.125. Bottom panel: Model G1.5.  
  The number of SNe Ia per year per unit
  stellar mass born in stars (at starburst $t=0$, 50\% binarity) is
  shown for the DDS (blue), SDS
  (red), HeRS (green), and
  sub-$M_{\rm Ch}$ (magenta) channels.  
  The sub-$M_{\rm Ch}$ SN Ia DTD
  clearly shows two distinct populations for Models A1 and G1.5: the
  helium star channel (spike at delay times ${\lesssim}
  500$~Myr) and the WD channel (from ${\sim} 800$~Myr to a
  Hubble time). The helium star channel however is absent in Model
  A.125. In the top panel we additionally show the
    DTD [SNuM] compiled by \citet[][table 1]{MSG10}, which is fit
    relatively well by a power-law $\sim t^{-1.2}$.  
    The data points showing the observed DTD are computed using 
    a different normalisation technique (see text), and thus we 
    show the points for comparison of the DTD shapes and not 
    the absolute numbers. We note that assuming a different binary
    fraction or IMF would change the level of our 
    normalisation.}
  \label{fig:dta1a125g1}
\end{figure}

In Figure~\ref{fig:dta1a125g1} we show the delay time distribution
of the four aforementioned progenitor channels for Models A1,
A.125 and Model G1.5.  We note that 
the bumpiness in the smoothed plot
is due to Monte Carlo noise. For our DTD normalisation of all models, we
have assumed a binary fraction across the entire initial stellar mass
function of $50$\% ($\frac{2}{3}$ of stars are in binaries), and  
we show the DTD normalised
to stellar mass (SNuM and SNe yr$^{-1}M_{\odot}^{-1}$). The 
mass represents the mass in formed stars, which
includes mass which has potentially been expelled from 
stars in SNe or thermal pulses for example.
In section 4.2, we give the delay times in tabular form.

Along with our theoretical DTDs, we show the observed 
(cosmic) DTD from the literature \citep{MSG10}. 
We wish only to compare the relative DTD shapes and not the 
absolute rates, since the normalisation of our {\tt StarTrack} 
DTD differs substantially from the normalisation 
techniques used in recovering the various observational DTDs.
The difference between the (higher) rates of observed SNe Ia 
and the rates from population synthesis is visible, and 
the apparent discrepancy is not yet fully resolved. It has been suggested 
that binary population synthesis codes tend to under-predict the  
SN Ia rates compared to the rates infered from recent 
observations, though one must 
keep in mind that many uncertainties are associated 
with the DTD recovery methods, i.e., extinction, star formation 
history, and the use of spectral population synthesis codes which 
neglect the existence of binaries (see \citet{DV04}, also \citet{ES09} 
have found that inclusion of massive binaries in spectral synthesis 
codes plays an important role in recovering accurate host galaxy properties).

\begin{figure}
  % \begin{center}
  \epsfig{file=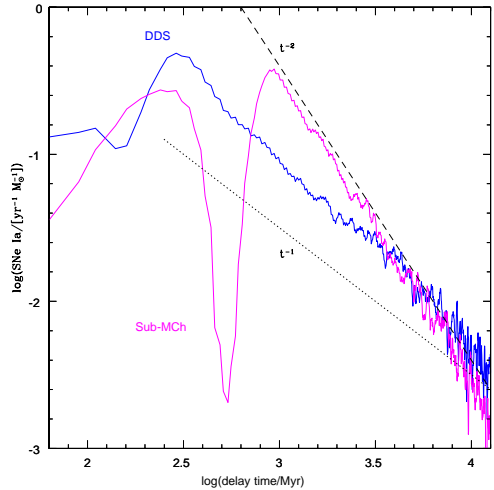, width=7cm,angle=0}\\
  \caption{ 
    Delay time distribution for the DDS (blue) and 
    sub-$M_{\rm Ch}$ (magenta) channels for model A1 (standard).  
    We show two power-laws alongside the DTDs: the DDS is relatively
    well-fit by a power-law $ t^{-1}$, where as the 
    sub-$M_{\rm Ch}$ model closely follows a power-law  
    with $t^{-2}$ beyond 1 Gyr, where all progenitors have helium-rich WD
    donors.}
  \label{fig:powerlaws}
  % \end{center}
\end{figure}

{\bf Model A1}. 
As was found in Paper~I Model 1, the DDS distribution
for Model A1 (top panel of Figure~\ref{fig:dta1a125g1}) follows a
power-law distribution with ${\sim} t^{-1}$ (see also
Figure~\ref{fig:powerlaws}), while the SDS
distribution is somewhat flat with no events with delay times less
than 460~Myr.  The reason why the SDS does not harbour very prompt
events is directly linked to the donor star's initial ZAMS mass.  When
the secondary ZAMS mass is ${>}2.8~M_{\odot}$, the binary will enter a
CE phase when the secondary fills its Roche-Lobe, rather than a stable
RLOF phase.  In such a case, the binary will not become an SDS SN Ia,
though may under the right circumstances evolve to SN Ia from the
HeRS channel.
The SDS events at long delay times originate from progenitors with 
very low-mass MS donors, which take many Gyr to evolve to contact 
under the influence of magnetic braking.
The HeRS DTD consists mostly of systems with relatively
short (${\sim} 100$~Myr--$2$~Gyr) delay times, with very few events at
longer delay times. We refer the reader to Paper~I for a description
of these DTDs.

The sub-$M_{\rm Ch}$ systems can easily by eye be grouped into two
classes: those prompt SNe which occur with delay times ${\lesssim}
500$~Myr, and those with delay times above ${\sim} 800$~Myr, with very
little overlap.  Not surprisingly, these two classes of SNe Ia stem
from two very different formation scenarios.  Those with short delay
times consist of progenitors which involve a helium-burning star
donor, whereas the rest mainly consists of helium WD donors (systems
with hybrid WD donors span ${\sim} 0.3$--$3$~Gyr delay times). 
We find that progenitors with helium star, helium WD and 
hybrid WD donors comprise 13\%, 78\% and 9\% of SNe Ia, 
respectively.  We note that ${\sim} 35$\% of sub-$M_{\rm Ch}$
SNe Ia explode within 1~Gyr of star formation.

The prompt component accounts for $13$\% of all sub-$M_{\rm
  Ch}$ SNe Ia that explode within 13~Gyr of star formation. Nearly all
of these systems (96\%) have helium star donors, with the 
rest having hybrid WD donors. The delay time is governed by the MS
lifetime of the donor star.  The companions with ZAMS masses
${\gtrsim} 3~M_{\odot}$ evolve off of the MS within ${\lesssim}
400$~Myr.  After the first CE, which leaves behind a CO primary WD and
a MS secondary star, the secondary (e.g., on the Hertzsprung gap) will
fill its Roche-Lobe and mass transfer is once again unstable leading
to a second CE phase.  The CE leaves the CO WD and newly-formed naked
helium star on a close orbit (${\sim} 35$--$40$~min).  Within a few
Myr, the orbit decreases to ${\sim} 25$~min, and the helium star fills
its Roche-Lobe.  However, initial mass transfer rates for the helium
star channel are low enough to enable
accumulation of the helium shell to commence immediately: typically
such systems have initial mass transfer rates ${\sim} 2 \times
10^{-8}~M_{\odot} \, \mathrm{yr}^{-1}$ \citep[for a discussion on the
evolution of low mass helium stars in accreting binaries
see][]{Yun08}.

The delayed component (delay times ${>} 500$~Myr) comprise the other
${\sim} 87$\% of the sub-$M_{\rm Ch}$ progenitors.  Binaries with
helium WD donors make up 90\% of the delayed component, 
while 10\% have hybrid donors.  These binaries also evolve
through two CE phases, as is expected for the evolution of AM CVn
binaries.  Similar to the DDS, the time-scale governing the DTD for
the helium WD channel is largely set by the gravitational radiation
time-scale \citep[see also][]{TY96}. However unlike the DDS, these WDs
do not merge upon contact, but enter a stable phase of RLOF.
Like the DDS DTD, the sub-$M_{\rm Ch}$ DTD follows 
a power-law above 1 Gyr, however with a steeper
functional form of $t^{-2}$ (Figure~\ref{fig:powerlaws}, see also 
section 5.2).

{\bf Model A.125.}  In the middle panel of
Figure~\ref{fig:dta1a125g1}, we show the DTD for Model A.125.
Contrary to the standard model, the DDS DTD is lacking progenitors at
longer delay times since on average the time a progenitor spends as a
detached double WD is decreased in this model (smaller orbital
separation following the CE phase).  For this model the SDS
progenitors can have shorter delay times compared to the standard
model due to the fact that the post-CE separations are overall
smaller.  Thus, a low CE efficiency model is more favourable for the
production of SDS SNe Ia.  The HeRS channel
has some very prompt events (helium star channel), although the low CE
efficiency serves to result in a merger during CE more frequently than
in Model A1.  The events at delay times ${\sim} 1$--$2$~Gyr belong to
progenitors with helium WDs, while events at long delay times ($>$ a
few Gyr) also involve helium WDs but belong to the evolved low-mass MS
donor channel discussed previously.  The DTD of the sub-$M_{\rm Ch}$
progenitors looks drastically different from that of the standard
model, and lacks a prompt component. This model is the only of the
three which does not display a prominent division of the sub-$M_{\rm
  Ch}$ progenitor channels; in fact there are no sub-$M_{\rm Ch}$ SNe
Ia originating from the helium star channel, since those progenitors 
will encounter unstable RLOF too early in their evolution.
With the adopted CE prescription, it is very difficult (or impossible)
to produce helium star donor channel sub-$M_{\rm Ch}$ SNe Ia within our model
framework, and thus there are no prompt events (the first SN Ia from
the sub-$M_{\rm Ch}$ channel occurs at ${\sim} 1.7$~Gyr).
 
{\bf Model G1.5.}  In Figure~\ref{fig:dta1a125g1}, bottom panel, we
show the DTD for Model G1.5.  Gravitational radiation plays a less
significant role for the DDS since following
the CE phase the binary orbit is still rather wide.  Similar to the
other two models, the DDS contributes the majority of its events at
very early times followed by a decline.  The SDS channel displays no
prompt events, because the first CE event does not lead to a dramatic
decrease in orbital separation. In general, the SNe Ia with short SDS
delay times from Model A1 will evolve into detached double CO WDs in
Model G1.5, since the binary orbit is not small enough for mass
transfer to begin once the secondary evolves off of the MS and fills
its Roche-Lobe.  The HeRS channel leads to
SNe Ia with very short delay times, though there are events at long
delay times but their frequency is for the most part too low to be
seen on the figure.  The sub-$M_{\rm Ch}$ DTD has the same general shape
as Model A1: the prompt and the delayed components.  We note however
that SNe Ia with delay times less than $1$~Gyr follow a different
evolutionary sequence compared to the corresponding events of Model
A1.  In Model A1, the first mass exchange interaction occurs when the
primary is an AGB star, where as for Model G1.5 the first mass
exchange event (CE or stable RLOF) occurs when the primary is less
evolved; a sub-giant or giant.  This occurs since the semi-latera recta
(and thus in general, the separations) of the G1.5 sub-$M_{\rm Ch}$
progenitors which explode in our models are smaller when mass 
transfer begins, as well as the fact
that the primaries for this model are somewhat more massive than
compared to the standard model and thus they evolve more quickly
%(compare ZAMS primary distributions of Figs. $1d$ and $3d$).

\subsection{Rates}
In Table 1, we show the DTDs in tabular form (rates as a function 
of epoch) for our models.
We estimate the Galactic SN Ia rate by convolving the DTD
  (in units of SNe/time) with a constant star formation history from 
  $0-10$ Gyr with a total mass born in stars of $6 \times
  10^{10}M_{\odot}$; see section 4 of Paper~I.  In this paper, we do not 
quantify the Galactic rate estimates explicitly since imposing a
particular star formation history serves to add sources of uncertainty 
to our DTD calculation, however we give some numbers as a guide.  In
principle, all of the important information is already presented in
the DTD plots and Table 1: it is possible to convolve the specific DTD 
with any star formation history of choice in order to achieve a
particular SN Ia rate for a 
given stellar population.

\begin{table}
  \centering
%  \begin{minipage}[H]{0.45\linewidth}
  \caption{Rates of SNe Ia (SNuM, 50\% binarity) for the four
    progenitor formation scenarios considered in this work, following a
    starburst at $t=0$.  Models A1 (left), A.125 (middle) and G1.5
    (right).}
  \begin{tabular}{l|c|c|c} %%{@{}lc@{}}
    \hline
    & A1   & A.125 & G1.5 \\
    \hline
    DDS  &   &  &  \\
    \hspace*{0.5cm} 0.1 Gyr & $2.0 \times 10^{-1}$ & $<10^{-4}$ & $2.0 \times 10^{-2}$ \\
    \hspace*{0.5cm} 0.5 Gyr & $1.6 \times 10^{-1}$ & $6.5 \times 10^{-2}$ &$2.2 \times 10^{-2}$\\
    \hspace*{0.5cm}  1 Gyr  & $8.0 \times 10^{-2}$ & $2.5 \times 10^{-2}$ & $5.3 \times 10^{-3}$\\
    \hspace*{0.5cm}  3 Gyr  & $2.5 \times 10^{-2}$ &  ${\lesssim}10^{-4}$ & ${\sim} 2 \times 10^{-3}$\\
    \hspace*{0.5cm}  5 Gyr  & $1.2 \times 10^{-2}$ & $0$  & ${\sim} 2 \times 10^{-3}$\\
    \hspace*{0.5cm} 10 Gyr  & ${\sim} 5 \times 10^{-3}$ & $0$ & ${\lesssim} 10^{-3}$\\ 
    \hspace*{0.5cm} & & &\\
    SDS             & & &\\
    \hspace*{0.5cm} 0.1 Gyr & $ 0 $ &  $10^{-3} $ & $ 0 $ \\
    \hspace*{0.5cm} 0.5 Gyr & ${\sim} 10^{-3}$  & $3.5 \times 10^{-3}$ & $ 0 $  \\
    \hspace*{0.5cm}  1 Gyr  & $1.5 \times 10^{-3}$ & $5 \times 10^{-3}$ & ${\lesssim} 10^{-3}$  \\
    \hspace*{0.5cm}  3 Gyr  & $2.0 \times 10^{-3}$ & ${\lesssim} 10^{-4}$ & ${\sim} 2 \times 10^{-3}$  \\ 
    \hspace*{0.5cm}  5 Gyr  & ${\sim} 1 \times 10^{-3}$ &${<}10^{-4}$  & ${\sim} 10^{-4}$  \\ 
    \hspace*{0.5cm} 10 Gyr  & ${\lesssim} 10^{-3}$ & ${\sim} 0$ & ${\lesssim} 10^{-4}$  \\
    \hspace*{0.5cm} & & & \\
    HeRS        & & &\\
    \hspace*{0.5cm} 0.1 Gyr & ${\sim} 3 \times 10^{-3}$ & $4 \times 10^{-3}$ & ${\lesssim} 10^{-3}$  \\
    \hspace*{0.5cm} 0.5 Gyr & $ 2.2 \times 10^{-2}$ & $0$  & $ {<} 10^{-3}$ \\
    \hspace*{0.5cm}  1 Gyr  & $ 8.0 \times 10^{-3}$ & ${<} 10^{-3}$ & $ {<} 10^{-4}$  \\
    \hspace*{0.5cm}  3 Gyr  & $ {<} 10^{-3}$ & ${\lesssim} 0$ & $ {<} 10^{-4}$   \\
    \hspace*{0.5cm}  5 Gyr  & $ {\lesssim} 10^{-4}$ & ${<} 10^{-4}$ & $ {<} 10^{-4}$ \\
    \hspace*{0.5cm} 10 Gyr  & $ {\sim} 0 $   & $ {\sim} 0 $ & $ {<} 10^{-4}$ \\
    \hspace*{0.5cm} & & & \\
    sub-$M_{\rm Ch}$ & & & \\
    \hspace*{0.5cm} 0.1 Gyr & ${\sim} 1 \times 10^{-1}$ & $0$ & ${\lesssim} 10^{-4}$  \\
    \hspace*{0.5cm} 0.5 Gyr & $ {\sim} 10^{-3}$ & $0$ & ${\sim} 10^{-3}$ \\
    \hspace*{0.5cm}  1 Gyr  & $ 3.3 \times 10^{-1}$ & ${<} 10^{-4}$ & ${\sim} 7 \times 10^{-2}$ \\
    \hspace*{0.5cm}  3 Gyr  & $ 4.0 \times 10^{-2}$ & ${<} 10^{-4}$ & ${\sim} 4 \times 10^{-3}$\\
    \hspace*{0.5cm}  5 Gyr  & $ 1.4 \times 10^{-2}$ & ${<} 10^{-3}$ & ${\sim} 2 \times 10^{-3}$\\
    \hspace*{0.5cm} 10 Gyr  & $ {\sim} 4 \times 10^{-3}$ & $\sim 0$ & ${\lesssim} 10^{-4} $ \\
    \hline
  \end{tabular}
%  \end{minipage}
\end{table}

\emph{Model A1}: This model produces the highest number of SNe Ia out
of our 3 models.  Table 1 (left) is very similar to table 1 (Elliptical
column) in Paper~I, though here for all tables we additionally include
the rates of sub-$M_{\rm Ch}$ SNe Ia, as well as two additional
epochs: 0.1 and 1~Gyr after star formation.  Slight variations between
the numbers in this study and table 1 of Paper~I are due to a slight
increase in volume of data, and thus a reduction in noise from
low-number statistics.  We find that the rate of our adopted
sub-$M_{\rm Ch}$ SN Ia model exceeds all other progenitor channels
between ${\sim} 0.7$ and $5$~Gyr, and these systems are enough to account for the
observed SN Ia rate, with a calculated Galactic rate of ${\sim} 2.6
\times 10^{-3}$ SN Ia yr$^{-1}$ (including all systems with a total WD
mass ${\gtrsim} 0.9~M_{\odot}$).  For comparison, the DDS rate is
${\sim} 2 \times 10^{-3}$ SN Ia yr$^{-1}$.  Both of these
values are within the estimate from \citet{CET99} of $4 \pm 2 \times
10^{-3} $ SN Ia yr$^{-1}$.  
As was determined in Paper~I, the Model A1
DDS rates are able to (just) account for the observed Galactic rate of
SNe Ia, whereas both the SDS and HeRS 
channels fall short by over an order of magnitude.

\emph{Model A.125}: This model produces the least SNe Ia progenitors
out of our three models.  The DDS is significantly decreased in number
(Table 1, middle), but is still the dominant channel at most times
under a few Gyr.  
The SDS rate exceeds the DDS rate above 3~Gyr, though the overall
rates are still too low for any progenitor in this model to account
the observed SN Ia rates.  
The rates of the HeRS SNe Ia are too low; many binaries do not survive both CE
events to become progenitors.  Similarly, 
the sub-$M_{\rm Ch}$ progenitors are not easily formed in this Model.

\emph{Model G1.5}: The overall rates for this model (Table 1, right) are
lower than found in the standard model, though not as low as found for
Model A.125.  In the DDS, since the binaries take a longer time to
reach contact (e.g., it can easily be more than a Hubble time), the
overall SN Ia rates are rather low compared to Model A1 with an
estimated Galactic rate of ${\sim} 2 \times 10^{-4}$~yr$^{-1}$, which
is about a factor of 10 too low.  The SDS channel produces very few
events before 2~Gyr, and matches those of the DDS at ${\sim} 3$~Gyr,
while the HeRS channel produces events with
delay times ${<}1$~Gyr and few events at later times.  Even
though the sub-$M_{\rm Ch}$ DTD exhibits the same general shape as
found in the standard model, the rates are overall too low being
roughly comparable to those of the DDS of this model (Galactic rate
estimate ${\sim} 3 \times 10^{-4}$~yr$^{-1}$).

\subsection{CO core masses}

\begin{figure}
\includegraphics[height=8in,clip]{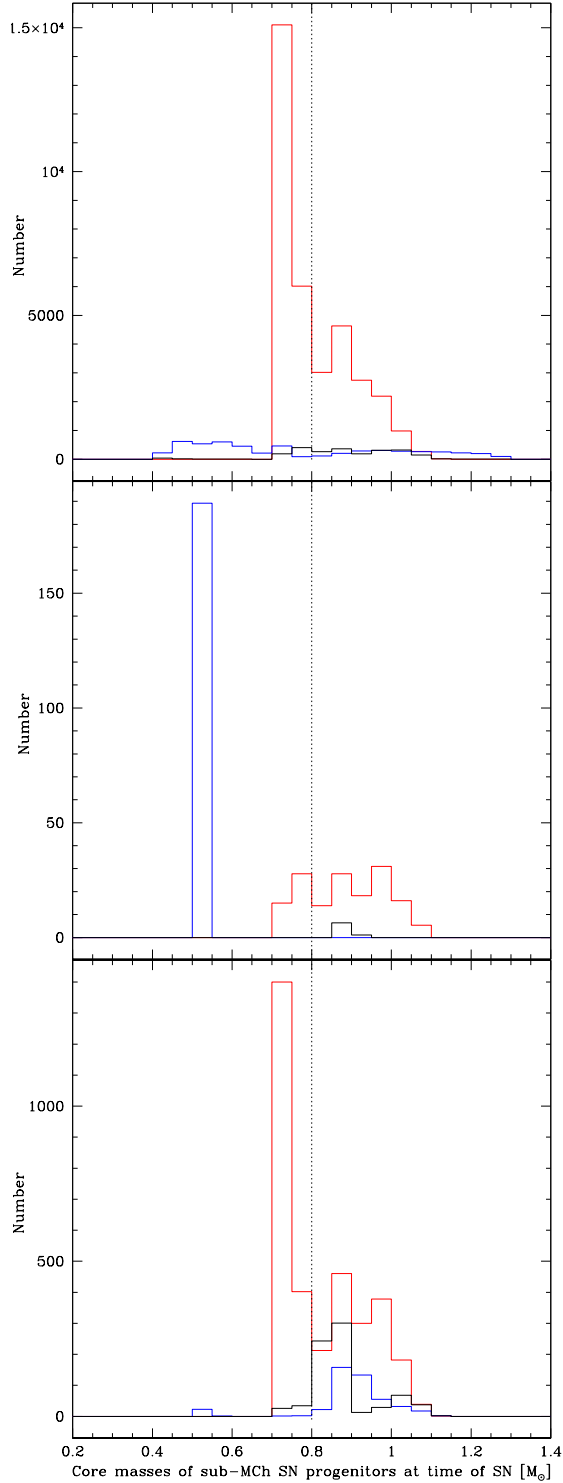}
  \caption{Distribution of {\tt StarTrack} CO WD core masses 
    which managed to accumulate a $0.1~M_{\odot}$ shell of
    helium. A double-detonation was assumed to follow in all cases.    
    A vertical line is drawn at M$_{\rm CO core}=0.8~M_{\odot}$, above
    which all systems are assumed to lead to sub-$M_{\rm Ch}$ SN Ia in
    our three models ($M_{\rm CO core} + M_{\rm He \, shell} =$ total WD
    mass). Top panel: Model A1. Middle panel: Model
    A.125.  Bottom panel: Model G1.5.
    The helium star channel is outlined in blue, the
    He-WD channel is outlined in red and the hybrid WD channel is
    outlined in black. Note different scales on the $y$-axes.}
  \label{fig:cores}
\end{figure}

In the sub-$M_{\rm Ch}$ scenario, the brightness is expected to be
largely determined by the mass of the underlying CO WD\@.  In
Figure~\ref{fig:cores}, we show the mass of the CO WD `core' (total WD
mass minus the helium shell mass) at time of SN Ia.  
As mentioned previously, 
a detonation of a ${\sim} 0.7~M_{\odot}$ core WD would
likely not look like a normal SN Ia. Since we currently lack a
theoretical lower mass limit for which exploding CO core masses could
potentially exhibit features which are characteristic of SNe Ia, for
completeness we show the CO core mass at explosion for the entire mass
spectrum for exploding sub-$M_{\rm Ch}$ cores.  We draw a
vertical line at $M_{\rm core} = 0.8~M_{\odot}$, above which the
systems are considered to be sub-$M_{\rm Ch}$ SNe Ia in our models.

The core mass distributions look very different for all three models.
In the top panel of Figure~\ref{fig:cores}, we show the core mass distribution for
Model A1. The progenitors of binaries with low core masses ($< 0.7~M_{\odot}$) go
through a different evolutionary channel than those with higher core
masses since they start out with smaller semi-latera recta and only
evolve through one CE event.  The cores associated with the
helium star channel span both low and high masses, 
though for our adopted sub-$M_{\rm Ch}$ scenario they have slightly
higher core masses on average compared to the WD channels.  The
hybrid WD channel shows a similarly flat distribution, which is not
unexpected since many of these systems undergo an evolutionary
sequence which is like that of a typical progenitor from the
helium star channel. For the helium WD channel which comprises the
majority, the masses decrease fairly steadily in number with
increasing mass, since there are simply a larger number of
less-massive CO WD cores to start with.  There is a clear lack of CO
core masses below $\sim~0.7 M_{\odot}$.  Typically these CO core
progenitors will accrete (and burn) at least $0.1~M_{\odot}$ (often
$\sim~0.2~M_{\odot}$) of helium at a high accretion rate before the
phase of helium accumulation begins for the $0.1~M_{\odot}$ shell, and
thus we find no CO cores from this channel with very low masses.
However, there are a number of exploding cores with masses 
$\sim~0.7-0.8~M_{\odot}$. One has to also consider the possibility that a
low-mass ($< 0.8~M_{\odot}$) CO core $+$ helium shell may not
reach sufficient conditions for a detonation to take place, which
might explain why we would not observe a large number of these 
events.

%%%look for ``as discussed in... 
For Model A.125 (Figure~\ref{fig:cores}, middle panel), the separation between the
helium star and double WD channels is quite distinct.  The lowest mass
CO cores belong to progenitors with helium star donors, and in our
adopted sub-$M_{\rm Ch}$ model all of these binaries have CO core
masses which are too low to qualify as SNe Ia.  The ZAMS masses of
these CO cores are small, ${\sim} 1.8$--$2.1~M_{\odot}$,
and these stars are unable to build a massive CO core before the first CE
is encountered.
Additionally, binaries which start their final RLOF phase when the
secondary is a helium star have lower initial accretion rates (${\sim}
10^{-8}~M_{\odot}$~yr$^{-1}$), which allows the CO core to immediately
accumulate (not burn) a shell of helium and produce a SN Ia without
the CO WD having to grow in mass by an extra
${\sim}0.1$--$0.2~M_{\odot}$.

The distribution of CO core mass for the helium WD channel of Model G1.5
(Figure~\ref{fig:cores}, bottom panel) is very similar to that of the standard
model.  However the different evolutionary sequences allowed in this
model enable the formation of more progenitors involving hybrid WD
donors.  The mass distributions for the helium star and hybrid WD
channels peak between $0.85$ and $0.9~M_{\odot}$ (total WD mass
0.95--1~$M_{\odot}$), which is a noteworthy feature, especially if
these systems are shown to contribute to the population of SNe Ia of
`normal' brightness \citep[][]{Sim10}.

\section{Discussion}

Recent hydrodynamic explosion simulations of sub-$M_{\rm Ch}$ CO WDs
\citep{Fin10} coupled with detailed nucleosynthesis and radiative
transfer modelling \citep{KS09} have revealed that sub-$M_{\rm Ch}$
mass SN Ia models exhibit features which are characteristically
similar to those observed in SNe Ia \citep{Sim10,Kro10}.  Motivated by
these new findings, as well as population synthesis rate estimates, we
have investigated sub-$M_{\rm Ch}$ SN Ia formation channels and have
calculated and presented the delay time distribution and rates of
their progenitors for three different parametrizations of the common
envelope phase.

We find that only the sub-$M_{\rm Ch}$ progenitor channel is able to
simultaneously
\begin{itemize}
\item {reproduce the observed rates for our standard model}
\item {provide an elegant explanation for the variety among SN Ia
    light-curves (mass of exploding WD)}
\item {naturally provide a system which is devoid of hydrogen}
\item {produce a DTD with distinct prompt (${\lesssim}
    500$~Myr)

    and delayed (${\gtrsim} 500$~Myr) components, originating from two
    channels with very different evolutionary time-scales}
\end{itemize}
We think that this last point is one of the most interesting,
considering the recent observational studies by different groups who
have found evidence for such a DTD \citep{Bra10,MB10,Mao10}.

\subsection{Double white dwarf mergers: implications}

We note both the works of \citet{Gui10}, who investigated detonations
in sub-$M_{\rm Ch}$ CO WDs undergoing rapid accretion during
dynamically unstable mass transfer from a helium-rich WD companion,
and \citet{van10}, who also considered mergers of WDs with a total
mass below $M_{\rm Ch}$ as possible progenitors of SNe Ia.  In this
study we do not investigate sub-$M_{\rm Ch}$ mergers in detail though
we briefly comment on them here.
We find that the number of sub-$M_{\rm Ch}$ WD mergers in our standard
model (considering all mergers where at least one WD is CO-rich, the
other being CO and/or helium-rich) is nearly twice that of DDS
mergers.  While it is generally believed that a WD merger with a total
mass below the Chandrasekhar mass limit would not lead to a SN Ia
explosion, these mergers should produce other interesting objects; R
Coronae Borealis stars are one example 
\citep[][see also \citet{BT09}]{Web84,Ibe96,Cla07} and these
types of merger events may be visible in upcoming transient surveys.
If we make a constraint similar to that of \citet{van10} counting both
sub-$M_{\rm Ch}$ and super-$M_{\rm Ch}$ WD mergers between CO WDs with
near-equal masses, we find that the number of mergers drops to 
${\sim} 42$\% of our standard model DDS rate, which is slightly
too low to explain all SNe Ia.

In our models we have assumed the commonly-adopted initial binary
orbital configurations for population synthesis studies: i.e., initial
separation flat in the logarithm (more binaries born on closer orbits
relative to large orbits), and thus the ZAMS distribution of all
semi-latera recta are the same for all three CE models.  However, we find that
for the low-CE efficiency case (Model A.125), DDS SNe Ia progenitors
are only formed from systems with initial (ZAMS) orbital
configurations which have rather large semi-latera recta compared to
those for our standard model.  In Model A.125, systems which would
have made DDS SNe Ia in Model A1 merge too early, and never make
double WDs.  It was already mentioned in \citet{HTP02} that
the initial distribution of orbital separations in population
synthesis studies should be distributed according to the (observed)
distribution of semi-latera recta rather than semi-major axes or
orbital periods alone.  We note here that an initial distribution
geared toward higher semi-latera recta than is canonically assumed
would serve to augment the
number of progenitors in models with low CE efficiency, making those DDS
rates closer to those of observations.

While the predicted rates of the DDS for our models do not conflict
with observations, these systems are theoretically expected to produce
neutron stars via AIC\@.  If this were the case, the AIC rate from the
AIC-merger channel alone would be ${\sim} 10^{-3}$ per year for the
Galaxy.  We find the {\tt StarTrack} AIC rate from the `RLOF-AIC'
channel is a factor of $10$ to $100$ less: no more than $10^{-4}$ per
year for our standard model.  This rate is in agreement with the upper
limit estimate derived from solar system abundances of neutron-rich
isotopes, which are expected to be produced in AICs
\citep{Fry99,Met09}.  However, if \emph{i}) population synthesis
estimates for the number of merging CO+CO WDs with a mass above
$M_{\rm Ch}$ are correct and \emph{ii}) in most environments these
mergers preferentially produce AICs and not SNe Ia, then this could
potentially be in conflict with the predicted abundance of
neutron-rich isotopes in the solar neighbourhood.  However, modelling
of AIC events, be it the `RLOF' or `merger' case, is still in its
infancy, and many uncertainties remain \citep[][see also
\citealt{Dar10}]{Des06,Des07,Met09}.  If one can say for certain that
AIC events formed from the merger of CO WDs produce very neutron-rich
ejecta, then this provides a potentially strong constraint on the
outcome of these mergers; namely that a non-negligible fraction of SNe
Ia must be formed through the DDS channel.  On the other hand, it is
possible that population synthesis calculations over-predict the number
of merging CO+CO WDs, which would also present an interesting problem
for the binary evolution community, and may challenge the idea that
the observed ${\sim} t^{-1}$ power-law DTD of SNe Ia originates from
double WD mergers alone.

\subsection{Further remarks on delay times}

The $t^{-1}$ power-law shape found in the delay time study of
\citet{Tot08} implies that the majority of progenitors in
elliptical-like galaxies originate from binaries for which the DTD is
most strongly governed by the time-scale associated with gravitational
wave radiation, thus these progenitors are likely to be DDS mergers
\citep[see section 3 of][]{RBF09}.  
However, we point out that a change in the CE removal efficiency $\alpha$ will 
have an effect on the amount of time between the last CE phase 
and final contact, thus affecting the shape (and perhaps to some degree the 
relevance) of the delayed DTD component.
Similar to the DDS, our study has
shown that the sub-$M_{\rm Ch}$ model DTD (e.g., Model A1) also
exhibits a power-law for delay times ${>} 1$~Gyr.  This is not
surprising, since the helium WD sub-$M_{\rm Ch}$ progenitors also
spend an appreciable time as detached compact stars evolving to
contact solely under the influence of gravitational radiation.
However, the DTD of the sub-$M_{\rm Ch}$ channel falls off more
steeply than the $t^{-1}$ power-law fit of \citet{Tot08} 
and the $t^{-1.1}-t^{-1.3}$ power-law fits of \citet{MSG10}, 
matching quite well to $t^{-2}$ (Figure~\ref{fig:powerlaws}).
Thus, when comparing our results to observationally-derived DTDs, the
DDS channel matches more closely than the sub-$M_{\rm Ch}$ channel.
However, a very recent study of Subaru/XMM-Newton Deep Survey (SXDS)
SNe Ia indicates that the DTD may be well-fit by a power-law of
$t^{-1.5}$ (J. Okumura, private communication 2010).  It is of course
possible that both DDS and sub-$M_{\rm Ch}$ progenitors contribute
substantially to the SN Ia population, potentially yielding a
DTD of functional form somewhere in between $t^{-1}$ and $t^{-2}$
above 1~Gyr, which would still be in agreement with the majority of
recent observations.

\subsection{Sub-Chandrasekhar SNe Ia connection to AM CVn stars and
  .Ia events}

Based on the observed local space density estimate of AM CVn
binaries\footnote{We would like to make the reader aware of the fact
  that population synthesis studies over-predict the number of AM CVn
  binaries in general compared to the observational
  results of \citet{RNG07} \citep[e.g.,][]{Nel01a,Rui10}. There
    may be several factors which conspire to cause the apparent 
    difference, though the various possibilities are not explored
    in this study.} performed
by \citet{RNG07}, \citet{Bil07} have calculated the occurrence rate of
the final (explosive) helium flash from ``.Ia'' systems in a
typical E/S0 galaxy with a mass of $10^{11}~M_{\odot}$ to be $(7 - 20)
\times 10^{-5}$~yr$^{-1}$; i.e., 2--6\% of the SN Ia rate in E/S0
galaxies.  .Ia events are expected to be about one tenth as
  bright as normal SNe Ia.  Since our sub-$M_{\rm Ch}$ progenitors
could also potentially lead to .Ia-like (and
not SN Ia) explosions, we think it is useful to independently estimate
the occurrence rate for such explosions in our standard (A1) model for
similar (E/S0 galaxy) conditions.
It was already found in section 4.2 that the sub-$M_{\rm Ch}$
rate assuming a burst of star formation at $t=0$ is ${\sim} 4 \times
10^{-3}$ at 10~Gyr.  Thus we find that among old stellar populations
our double-detonation thermonucler explosions will be roughly 
$30$ times more frequent than the estimated .Ia explosion rate of 
\citet{Bil07}.

We also note that in the study of \citet{Bil07} it was found that the
ignition mass of the helium shell in .Ias varies as a function of the
underlying CO core mass and the rate of accretion.  However for our
first investigation of double-detonation sub-$M_{\rm Ch}$ SNe we have
used a more simplified model in which the ignition mass is always the
same ($0.1~M_{\odot}$).  The consequences of this on the resulting SN
rate are not expected to be too drastic, as the time-scale for the
helium accretion is relatively short compared to the evolutionary
lifetime of the progenitors.  This is particularly true for the
helium WD donor case, which is the scenario most relevant for
\citet{Bil07}.

\citet{YL04} found that rotation may pose a problem for the 
initiation of a detonation in accreted helium shells.  They found 
that the spin-up of the WD due to the accretion and resulting 
dissipation due to differential rotation might cause helium flashes 
to occur for lower shell masses, which may lead to inhibition of a 
detonation in turn resulting in fewer sub-$M_{\rm Ch}$ SNe. 
However, \citet{PB04} found that the accreted material will be 
brought into co-rotation with the WD already at low depths within 
the helium shell, and so as noted in \citet{Bil07} rotation should 
not play a significant role in the heating of the helium shell and 
subsequent helium-ignition. 

\subsection{The link between sub-Chandrasekhar SNe Ia and their
  progenitors}

While it is useful to understand how the host galaxy environment
influences the SN ejecta/observables, it is also fundamentally
important to find a direct physical connection between the progenitor
population and the observational characteristics of SNe Ia.  For some
time it has been known that brighter SNe Ia occur more frequently
among young stellar populations \citep{Ham95}.  Could it be possible
that sub-$M_{\rm Ch}$ SNe Ia arising from the (prompt) helium star
channel are brighter than those from the double WD channel?  This may
be the case particularly considering Model~A1, where the core mass of
the exploding star for the helium star channel is on average slightly
larger than for the double-WD channel (see Figure~\ref{fig:cores}), and thus is
likely to produce more $^{56}$Ni.  We also note that for both Model A1
and Model G1.5, ${\sim} 70$\% of progenitors with delay times ${<}
1$~Gyr have CO WD masses ${>} 1.0~M_{\odot}$ (CO core masses ${>}
0.9~M_{\odot}$), while this fraction is only ${\sim} 45$--$50$\% for
progenitors with delay times ${>} 3$~Gyr.  However there is no
\emph{strong} trend in our models which indicates that more massive
WDs explode among younger populations.  The majority of the
sub-$M_{\rm Ch}$ binaries are double WDs, and the MS lifetime (ZAMS
mass) of the primary star does not play a dominant role in setting the
delay time.

Another point worth considering is that the helium star channel
progenitors undergo two CE events on a relatively short time-scale
compared to the time the stars spend as a post-MS detached binary.
Thus these binary systems should be hotter and may be more readily
detectable than their (colder, longer lived) double-WD counterparts.
Since these helium star channel SNe in our models are expected to occur a few Myr
after the last CE phase, the detection of such an explosion will
probably not be inhibited by circumstellar matter from the companion.
However, since these explosions involving helium stars are expected to
be found among young stellar populations, they are likely to occur in
regions of active star formation where their detection may be thwarted
by the presence of dust and possibly circumstellar matter from nearby
stellar systems.  The binary progenitors of the helium WD channel on
the other hand, although more abundant at most delay times, should be
harder to detect as most of their evolutionary time is spent during
the detached double WD phase.

Thus far, we have only found (possibly) a weak correlation between the
mass of the exploding WD and delay time, making it difficult to infer
a connection between observed brightness ($^{56}$Ni synthesised in the
explosion) and progenitor age.  Nevertheless, If a connection between
the age of the primary CO WD and the production of $^{56}$Ni can be made
in sub-$M_{\rm Ch}$ explosions such that dimmer SNe Ia occur
among older populations, this would have very exciting consequences
for our study.

\subsection{Conclusion}

Our standard model population synthesis indicates that there are
potentially enough sub-$M_{\rm Ch}$ progenitors to account for the
rates of SNe Ia. Nevertheless, much uncertainty still remains regarding the
formation and evolution of close binary stars: mass transfer and
accretion efficiencies, effects of rotation and magnetic fields,
impact of metallicity on stellar winds and subsequent stellar and
binary evolution, the common envelope phase, etc.  Even given a large
population of potential progenitors for sub-$M_{\rm Ch}$ explosions,
there remain open questions about the explosion itself. Hydrodynamical
studies have previously shown that sub-$M_{\rm Ch}$ WDs with an
overlying helium shell can undergo a double-detonation which looks
like a SN Ia, though the real answer as to what fraction of these
systems lead to SNe Ia explosions depends on specific details.  Most
critically, under exactly which conditions does helium ignition occur,
and how does the nucleosynthesis proceed?
 
The sub-$M_{\rm Ch}$ model is the first model which demonstrates 
a sufficient number of SNe Ia events to account for all, or at least
some substantial fraction of, SNe Ia (Model~A1), as well as \emph{two
  distinct formation channels with their own characteristic DTD}: A
prompt (${<}500$~Myr) helium star channel originating from binaries
with more massive secondaries, and a more delayed (${>}500$~Myr)
double WD channel originating from AM CVn-like progenitor binaries
with lower mass.  Whether some or all of the sub-$M_{\rm Ch}$ models
explored in this work really lead to thermonuclear explosions that
look like normal (or some subclass of) SNe Ia is still a topic
which requires further study.

\section*{Acknowledgments}
The authors thank the anonymous referee for helpful comments and
suggestions.  
AJR would like to thank the organisers of the 2010 Workshop on Nuclear
Astrophysics at Ringberg Castle, the organisers of the Workshop on
Progenitors of SNe Ia at the Lorentz Center in Leiden, as well as the
following people in general for helpful discussions: B. Metzger,
I. Seitenzahl, L. Bildsten, K. Shen, T. Janka, B. Wang, C.
Heinke, E. Van den Heuvel, R. Pakmor, J. Okumura and P. Mazzali.  Also, AJR  
thanks J.~Grindlay and the ChaMPlane group at the
Harvard-Smithsonian Center for Astrophysics for use of computing
resources, where the {\tt StarTrack} simulations were performed.

\label{lastpage}

\end{document}